\documentclass[letterpaper, 10 pt, conference]{ieeeconf}  

\IEEEoverridecommandlockouts                              
\title{\LARGE \bf
Towards improving the estimation performance of a given nonlinear observer: a multi-observer approach
}

\author{E. Petri, 	R. Postoyan, D. Astolfi, D. Ne\v{s}i\'c and V. Andrieu
	\thanks{
		E. Petri and R. Postoyan are with the Universit\'e de Lorraine, CNRS, CRAN, F-54000 Nancy, France.   
		(\texttt{elena.petri@univ-lorraine.fr}, \texttt{romain.postoyan@univ-lorraine.fr}).}
	\thanks{ D. Ne\v{s}i\'c is with the Department of Electrical and Electronic Engineering, The University of Melbourne, Parkville, 3010 Victoria Australia.
		(\texttt{dnesic@unimelb.edu.au}). }
	\thanks{D. Astolfi and V. Andrieu are with  
		Universit\'e Claude Bernard Lyon 1, CNRS, LAGEPP UMR 5007, F-69100,
		Villeurbanne, France. (\texttt{daniele.astolfi@univ-lyon1.fr}, \texttt{vincent.andrieu@univ-lyon1.fr}).}
	\thanks{This work was funded by Lorraine Universit\'e d'Excellence LUE, HANDY project ANR-18-CE40-0010-02, the France Australian collaboration project IRP-ARS CNRS and the Australian Research Council under the Discovery Project DP200101303.  }
}

\usepackage{graphicx,fleqn,amstext,amsmath,amssymb,color,psfrag,mathabx,booktabs}
\usepackage[latin1]{inputenc}
\usepackage[colorlinks=true, letterpaper,    
urlcolor=black,     
citecolor=blue,
menucolor=black,    
linkcolor=black,    
pagecolor=black,    
breaklinks=true,
]{hyperref}

\usepackage{amsthm}

\usepackage[noadjust]{cite}

\usepackage{comment}

\newcommand{\R}{\ensuremath{\mathbb{R}}}

\newcommand{\Rlo}{\ensuremath{\mathbb{R}_{\geq 0}}}

\newcommand{\Zo}{\ensuremath{\mathbb{Z}_{\geq 0}}}
\newcommand{\Zp}{\ensuremath{\mathbb{Z}_{> 0}}}
\newcommand{\Z}{\ensuremath{\mathbb{Z}}}

\definecolor{bleucit}{rgb}{0.2,0.4,0.6} 

\definecolor{blue_cv}{rgb}{0.09,0.35,0.78}

\newcommand{\KL}{\ensuremath{\mathcal{KL}}}
\newcommand{\K}{\ensuremath{\mathcal{K}}}
\newcommand{\Kinf}{\ensuremath{\mathcal{K}_{\infty}}}

\newcommand{\argmin}{\ensuremath{\text{argmin}\,}}

\newcommand{\dom}{\ensuremath{\text{dom}\,}}

\newcommand{\sat}{\ensuremath{\text{sat}}}

\newcommand{\norm}[1]{\ensuremath{\left\|{#1}\right\|}}

\theoremstyle{theorem}

\newtheorem{ass}{\textnormal{\textbf{Assumption}}}
\newtheorem{prop}{Proposition}

\newtheorem{lem}{Lemma}

\newtheorem{thm}{\textnormal{\textbf{Theorem}}}

\newtheorem{rem}{\textnormal{\textbf{Remark}}}

\usepackage{array}

\usepackage{mathtools}

\usepackage[noadjust]{cite}

\let\labelindent\relax
\usepackage{enumitem}
\usepackage{tikz}
\usetikzlibrary{shapes,arrows,calc}
\definecolor{MyGreen}{RGB}{50,140,80}

\usepackage[export]{adjustbox}
\usepackage{ circuitikz }
 \usetikzlibrary{arrows}
 
\usepackage{algorithm}
\usepackage{algpseudocode}

\usetikzlibrary{backgrounds}

\begin{document}

\maketitle
\thispagestyle{empty}
\pagestyle{empty}

\begin{abstract}
Various methods are nowadays available to design observers for broad classes of systems. Nevertheless, the question of the tuning of the observer to achieve satisfactory estimation performance remains largely open. This paper presents a general supervisory design framework for online tuning of the observer gains with the aim of achieving various trade-offs between robustness and speed of convergence.
We assume that a robust nominal observer has been designed for a general nonlinear system and the goal is to improve its performance. We present for this purpose a novel hybrid multi-observer, which consists of the nominal one and a bank of additional observer-like systems, that are collectively referred to as modes and that differ from the nominal observer only in their output injection gains. We then evaluate on-line the estimation cost of each mode of the multi-observer and, based on these costs, we select one of them at each time instant.
%
Two different strategies are proposed. In the first one, initial conditions of the modes are reset each time the algorithm switches between different modes. In the second one, the initial conditions are not reset.
We prove a convergence property for the hybrid estimation scheme and we illustrate the efficiency of the approach in improving the performance of a given nominal high-gain observer on a numerical example.
%
\end{abstract}

\section{Introduction}
State estimation of dynamical systems is a central theme in control theory, whereby an observer is designed to estimate the unmeasured system states exploiting the knowledge of the system model and input and output measurements. 
Many techniques are available in the literature for the observer design of linear and nonlinear systems, see \cite{bernard2022observer} and the references therein. 
The vast majority of these works concentrate on designing the observer so that the origin of the associated estimation error system is (robustly) asymptotically stable. 
A critical and largely open question of how to tune the observer to obtain desirable properties 
in terms of convergence rate, overshoot and robustness to model uncertainties, measurement noise and disturbances remains. 
 One of the challenges is that there are different trade-offs between these properties. 
An answer to this question in the special context of linear systems affected by additive Gaussian noise impacting the dynamics and the output is the celebrated Kalman filter \cite{kalman1961new}. For general nonlinear systems on the other hand, it is very hard to design an optimal observer, as this requires solving challenging partial differential equations \cite{helton1999extending}. In this context, an alternative consists in aiming to improve the estimation performance of a given observer. 
To the best of the authors' knowledge, existing works in this direction either concentrate on specific classes of systems, see e.g., \cite{li2015finite, rios2018hybrid, alessandri2022hysteresis} for linear systems and e.g., \cite{astolfi2015high, astolfi2018low, esfandiari2019bank, bernat2015multi, mousavi2021low, ahrens2009high} in the context of high-gain observers, or on specific properties like robustness to measurement noise, see, e.g.,  \cite{astolfi2021stubborn, battilotti2021performance}.
%
An exception is \cite{astolfi2019uniting}, where two observers designed for a general nonlinear system are ``united'' to exploit the good properties of each of them. \textcolor{black}{However, the design in \cite{astolfi2019uniting} is not always easy to implement as it requires knowledge of various properties of the observers (basin of attraction, ultimate bound), which may be difficult to obtain.}

This paper presents a flexible and general observer design methodology based on supervisory multi-observer ideas that can be used to address various trade-offs between robustness to modeling errors and measurement noise and convergence speed.
%
A multi-observer consists of a bank of observers that run in parallel. It has been used in the literature in a range of different contexts. For instance, for high-gain observers, it was shown in e.g., \cite{esfandiari2019bank, bernat2015multi, mousavi2021low} that a multi-observer can improve the sensitivity to measurement noise and/or reduce the undesired overshoot during the transient, also known as peaking phenomenon. 
In \cite{chong2015parameter, aguiar2008identification, aguiar2007convergence}, multi-observers have been proposed for the joint estimation of system states and unknown parameters. 
 %
 
In this paper we propose a new problem formulation that we believe has not yet been addressed in the literature. In particular,
%
we present a novel multi-observer scheme with the aim of improving the estimation performance of a given nominal observer designed for a general nonlinear system. 
The nominal observer is assumed to be such that the associated estimation error system satisfies an input-to-state stability property 
with respect to measurement noise and disturbances. 
A broad range of nonlinear observers in the literature satisfy this property,
see \cite{astolfi2021stubborn, shim2015nonlinear} and the references therein. We then construct a multi-observer, composed of the nominal observer and additional dynamical systems, that have the same form as the nominal one, but with different gains. It is important to emphasize that the gains can 
 be arbitrarily assigned. Consequently, this design flexibility can be used to address a range of very different design trade-offs between robustness and convergence speed. 
Because the gains are different, each element of the multi-observer, referred to as 
mode, exhibits different properties in terms of speed of convergence and robustness to measurement noise. 
We run all modes in parallel and we evaluate their estimation performance using monitoring variables, inspired by supervisory control and estimation techniques, see e.g. \cite{chong2015parameter, hespanha2003hysteresis, morse1996supervisory, hespanha2001multiple}. 
Based on these monitoring variables, we design a switching rule that selects one mode at any time instant. The modes that are not selected at a switching instant are either unchanged or their state estimates, as well as their monitoring variables, are reset to the ones of the selected mode. 

We model the overall system as a hybrid system using the formalism of \cite{goebel2012hybrid}. 
We prove that the proposed hybrid estimation scheme satisfies an input-to-state stability property with respect to deterministic disturbance and measurement noise. 
	The result of this paper provides a flexible and general framework for addressing the above mentioned trade-offs of the state estimation of nonlinear systems.
	We show that the performance of the proposed hybrid multi-observer  is, at least, as good as the performance of the nominal observer by design. We also illustrate on a numerical example that the proposed technique may significantly improve the estimation performance compared with the nominal observer.  


The paper is organized as follows. 
Preliminaries are presented in Section~\ref{Notation} and 
the problem statement is given in Section~\ref{ProblemStatement}. Section~\ref{HybridEstimationScheme} presents the proposed hybrid estimation scheme and we model the overall system as a hybrid system in Section~\ref{HybridModel}. The stability of the hybrid system is analyzed in Section~\ref{Main result}. 
A numerical case study on a high-gain observer for a Van der Pol oscillator is reported in  Section~\ref{Example}.
Finally, Section~\ref{Conclusions} concludes the paper. 
%
Proofs are given in the Appendix, some of them are omitted for space reasons.
\color{black}
\section{Preliminaries}\label{Notation} 
The notation $\R$ stands for the set of real numbers and $\Rlo:= [0, +\infty)$. 
We use $\Z$ to denote the set of integers, $\Zo:= \{0,1,2,...\}$ and $\Zp:= \{1,2,...\}$. For a vector $x \in \R^n$, $|x|$ denotes its Euclidean norm. For a matrix $A \in \R^{n  \times m}, \norm{A}$ stands for its 2-induced norm. For a signal $v: \R_{\geq0} \to \R^{n_v}$, with $n_v \in \Zp$, $\norm{v}_{[t_1, t_2]}:= \textnormal{ess} \sup_{t \in [t_1, t_2]} |v(t)|$. 
Given a real, symmetric matrix $P$, its maximum and minimum eigenvalues are denoted by $\lambda_{\max}(P)$ and $\lambda_{\min}(P)$ respectively. The notation $I_N$ stands for the identity matrix of dimension $N \in \Zp$. 
%
%
We consider $\K_\infty$ and $\KL$ functions as defined in \cite[Definitions 3.4 and 3.38]{goebel2012hybrid}.
%
%
%
Given a function $f: \mathcal{S}_1\to\mathcal{S}_2$ with sets $\mathcal{S}_1$, $\mathcal{S}_2$, $\dom f:=\{z\in\mathcal{S}_1 : f(z)\neq\emptyset\}$.
%
Based on the formalism of \cite{goebel2012hybrid}, we model the proposed estimation scheme together with the plant as a hybrid system with inputs of the form
\begin{equation}
	\mathcal{H} \;:\; \left\{
	\begin{array}{rcll}
		\dot x &=& F(x,u), & \quad x \in \mathcal{C}, 
		\\
		x^+ & \in & G(x,u),  &\quad x \in \mathcal{D},
	\end{array}
	\right.
	\label{eq:hybridSystemNotation}
\end{equation}
where 
$\mathcal{C}\subseteq \R^{n_x} $ is the flow set, 
$\mathcal{D}\subseteq \R^{n_x}$ is the jump set,
$F$ is the flow map and $G$ is the jump map. 
We consider hybrid time domains as defined in \cite{goebel2012hybrid}.
We use the notion of solution for system \eqref{eq:hybridSystemNotation} as given in \cite[Definition 4]{heemels2021hybrid}.
Given a set $\mathcal{U} \subseteq \R^{n_u}$, $\mathcal{L}_{\mathcal{U}}$ is the set of all functions from $\R_{\geq0}$ to $\mathcal{U}$ that are Lebesgue measurable and locally essentially bounded. 
%
Given a set $\mathcal{C} \subseteq \R^n$, the tangent cone to the set $\mathcal{C}$ at a point $x \in \R^n$, denoted $T_\mathcal{C}(x)$, is the set of all vectors $v \in \R^n$ for which there exist $x_i \in \mathcal{C}$, $\tau_i > 0$, with $x_i \to x$, $\tau_i \to 0$, and $v = \lim_{i \to \infty} \frac{x_i - x}{\tau_i}$.
Finally, we use $U^\circ(x,v):= \limsup_{h \to 0^+, y \to x}\frac{U(y+hv) - U(y)}{h}$ to denote the Clarke generalized directional derivative of a Lipschitz function $U$ at $x$ in the direction $v$ 
 \cite{clarke1990optimization}.
%

\section{Problem statement}\label{ProblemStatement}
The aim of this work is to improve the estimation performance of a given nonlinear nominal observer by exploiting a novel hybrid estimation scheme that is presented in Section~\ref{HybridEstimationScheme}. 
%
We consider the plant model
\begin{equation}
	\begin{aligned}
		\dot x  &=  f_p(x,u,v)\\
		y  &=  h(x, w),
	\end{aligned}
	\label{eq:system}
\end{equation}
where $x \in \R^{n_x}$ is the state to be estimated, $u \in \R^{n_u}$ is the measured input, 
$y \in \R^{n_y}$ is the measured output, $v \in \R^{n_v}$ is an unknown disturbance input and $w \in \R^{n_w}$ is an unknown measurement noise, with $n_x, n_y \in \Zp$ and $n_u, n_v, n_w \in \Zo$. The input signal $u: \R_{\geq 0} \rightarrow \R^{n_u}$, the unknown disturbance input $v:\R_{\geq 0} \rightarrow \R^{n_v}$ and the measurement noise $w:\R_{\geq 0} \rightarrow \R^{n_w}$ are such that $u \in \mathcal{L_U}$, $v \in \mathcal{L_V}$ and $w \in \mathcal{L_W}$ for closed sets $\mathcal{U} \subseteq \R^{n_u}$,  $\mathcal{V} \subseteq \R^{n_v}$ and $\mathcal{W} \subseteq \R^{n_w}$. 
%
We consider a so-called nominal observer for system~\eqref{eq:system} of the form
\begin{equation}
	\begin{aligned}
		\dot{\hat{x}}_{1}  &=  f_o(\hat{x}_{1},u, L_{1} (y-\hat{y}_{1}))\\
		\hat{y}_{1}  &=  h(\hat{x}_{1}, 0),
	\end{aligned}
	\label{eq:observerNominal}
\end{equation}
where $\hat{x}_{1}\in \R^{n_x}$ is the state estimate, $\hat{y}_{1} \in \R^{n_y}$ is the output estimate and $L_{1} \in \R^{n_{L_1} \times{n_y}}$ is the observer output injection gain with $n_{L_1} \in \Zp$.
 We define the estimation error as $e_{1} := x -\hat{x}_{1} \in \R^{n_x}$ and introduce a \textit{perturbed} version of the error dynamics as, in view of \eqref{eq:system} and \eqref{eq:observerNominal}, 
 \begin{equation}
 	\begin{aligned}
 		\dot e_1 &= f_p(x,u,v,w)- f_o(\hat{x}_{1},u, L_{1}(y - \hat{y}_{1}) +d )\\
 		&=:\tilde f(e_1,x,u, v, w, d)
 	\end{aligned}
 \label{eq:PerturbedNominalEstimationError}
 \end{equation}
where $d \in \R^{n_{L_1}}$ represents an additional perturbation. 
 We assume that observer \eqref{eq:observerNominal} is designed such that 
system \eqref{eq:PerturbedNominalEstimationError} is input-to-state stable with respect to $v$, $w$ and $d$, uniformly with respect to $u$, as formalized next. 
\begin{ass}
	There exist $\underline{\alpha}$, $\overline{\alpha},  \psi_1, \psi_2 \in \Kinf$, 
	$\alpha \in \R_{> 0}$, $\gamma\in \R_{\geq 0}$
	and  $V: \R^{n_x} \rightarrow \R_{\geq0}$ continuously differentiable, such that for all $x \in \R^{n_x}$, $\hat{x}_{1} \in \R^{n_x}$, $d \in \R^{n_{L_1}}$, $u \in \mathcal{U}$,  $v \in  \mathcal{V}$, $w \in \mathcal{W}$, 
	\begin{equation}
		\underline{\alpha}(|e_{1}|) \leq V(e_{1}) \leq \overline{\alpha}(|e_{1}|)	
		\label{eq:NominalAssumptionSandwichBound}	
	\end{equation}
	\begin{equation}
	\begin{array}{r}
		\hspace{-0.5em} \left\langle \nabla V(e_{1}), \tilde f(e_1, x,u, v, w, d) \right\rangle  
		\leq 
		-\alpha V(e_{1}) + \psi_1(|v|) \\[0.5em] 
		+ \psi_2(|w|)  + \gamma|d|^2,
	\end{array}
	\label{eq:NominalAssumptionDerivative}
\end{equation}	
with $e_1=x-\hat x_1$.
	\hfill $\Box$
	\label{NominalAssumption}
\end{ass}
A large number of observers in the literature have the form of \eqref{eq:observerNominal} and satisfy Assumption~\ref{NominalAssumption}, possibly after a change of coordinates, see \cite{astolfi2021constrained, astolfi2021stubborn, shim2015nonlinear} and the references therein for more details. 
 Assumption~\ref{NominalAssumption} implies that 
 there exist $\beta \in \KL$ and $\rho \in \Kinf$ such that, for any initial conditions $x(0)$ and $e_1(0)$ to \eqref{eq:system} and \eqref{eq:PerturbedNominalEstimationError} with $u \in\mathcal{L}_{\mathcal{U}}$, $v \in \mathcal{L_V}$, $w \in \mathcal{L_W}$ and $d \in \mathcal{L}_{\R^{n_{L_1}}}$, the corresponding solution  verifies, for all $t\in \dom x \cap \dom e_1$,  
 \begin{equation}
 	|e_{1} (t)| \leq \beta(|e_{1}(0)|, t) + \rho(\norm{v}_{[0,t]} +\norm{w}_{[0,t]} + \norm{d}_{[0,t]}).
 	\label{eq:ISSnominalConverging}
 \end{equation}
Equation 
\eqref{eq:ISSnominalConverging} provides a desirable robust stability property of the estimation error associated to observer \eqref{eq:observerNominal}. However, this property may not be fully satisfactory in terms of performance, like convergence speed and noise/disturbance rejection. 
%
%
Our objective is therefore to propose a hybrid redesign of observer \eqref{eq:observerNominal}, which aims at improving its performance, 
 in a sense made precise later, while still preserving an input-to-state stability property for the obtained estimation error system.
%
For this purpose we propose a novel hybrid multi-observer scheme, which is presented in the next section.

%
\section{Hybrid estimation scheme}\label{HybridEstimationScheme}
 \begin{figure*}[]
	\begin{center}
		\tikzstyle{blockB} = [draw, fill=blue!20, rectangle, 
		minimum height=3.5em, minimum width=7em]  
		\tikzstyle{blockG} = [draw, fill=MyGreen!20, rectangle, 
		minimum height=3.5em, minimum width=8.5em]
		\tikzstyle{blockR} = [draw, fill=red!40, rectangle, 
		minimum height=2em, minimum width=3em]
		\tikzstyle{blockO} = [draw,minimum height=1.5em, fill=orange!20, minimum width=4em]
		\tikzstyle{input} = [coordinate]
		\tikzstyle{blockW} = [draw,minimum height=3.5em, fill=white!20, minimum width=6em]
		\tikzstyle{blockWplant} = [draw,minimum height=1.5em, fill=white!20, minimum width=4em]
		\tikzstyle{input1} = [coordinate]
		\tikzstyle{blockCircle} = [draw, circle]
		\tikzstyle{sum} = [draw, circle, minimum size=.3cm]
		\tikzstyle{blockSensor} = [draw, fill=white!20, draw= blue!80, line width= 0.8mm, minimum height=10em, minimum width=13em]
		\tikzstyle{blockDOT} = [draw,minimum height=8em, fill=white!20, minimum width=14em, dashed]
		\tikzstyle{blockMuxGreen} = [draw, minimum height=21em, fill=MyGreen!30, line width=0.1mm]
		\tikzstyle{blockMuxBlueBig} = [draw, minimum height=21em, fill=blue!30, line width=0.1mm]
		\tikzstyle{blockMuxBlueSmall} = [draw,minimum height=6.5em, fill=blue!30, minimum width=0.8mm]
		
		\begin{tikzpicture}[auto, node distance=2cm,>=latex , scale=0.82,transform shape] 
			
			\node [input, name=input] {};
			\node [input, right of= input, node distance=0.6cm] (inputLine) {};
			\node [blockO, right of=input, node distance=1.7cm] (plant) {   
				$\begin{array}{c}
					\textbf{Plant} \\ (x)
				\end{array}$
			};
			
			\draw [draw,->] (input) -- node {} (plant);
			\draw [draw,-] (input) -- node [pos=0.3]{\large $u$} (plant);
			\node [input, above of= plant, node distance=1.3cm] (disturbance) {};
			\draw [draw,->] (disturbance) -- node [pos=0.5]{\large $v$, $w$} (plant);
			
			\node [input, right of= plant, node distance=2cm] (output) {};
			\draw [draw,-] (plant) -- node  [pos=0.5]{\large $y$} (output);
			\node [input, above of= output, node distance=3cm] (NominalObserverLine) {};
			\draw [draw,-] (output) -- (NominalObserverLine);
			\node [blockB, right of=NominalObserverLine, node distance=2.5cm] (NominalObserver) {
				$\begin{array}{c}
				\textbf{Nominal Observer}\\
				\textbf{Mode $1$}
				\end{array}$
			};
			\draw [draw,->] (NominalObserverLine) -- (NominalObserver);
			
			\node [input, above of= output, node distance=1cm] (Observer2Line) {};
			\node [blockB, right of=Observer2Line, node distance=2.5cm] (Observer2) {
				\textbf{Mode $ {2}$}
			};
			\draw [draw,->] (Observer2Line) -- (Observer2);
			
			\node [input, below of= output, node distance=4cm] (ObserverN1Line){}; 
			\draw [draw,-] (output) -- (ObserverN1Line);
			\node [blockB, right of=ObserverN1Line, node distance=2.5cm] (ObserverN1) {
				\textbf{Mode $ {N+1}$}
			};
			\draw [draw,->] (ObserverN1Line) -- (ObserverN1);
			\node at ($(Observer2)!.45!(ObserverN1)$) {\large \textcolor{blue!80}{\vdots}};
			
			\node [input, above of= NominalObserver, right = 1.55cm, node distance=0.25cm] (StateEstimate1line) {};
			\node [input, above of= NominalObserver, left = 1.70cm, node distance=0.25cm] (NominalObserverInput) {};
			\node [input, left of= NominalObserverInput, node distance=1cm] (NominalObserverInputStart) {};
			\node [input, above of= inputLine, node distance=3.25cm] (NominalObserverInputLeft) {};
			\draw [draw,-](inputLine) -- node  {}(NominalObserverInputLeft);
			\draw [draw,-](NominalObserverInputLeft) -- node  {}(NominalObserverInputStart);
			\draw [draw,->](NominalObserverInputStart) -- node  [pos=0.5]{\large \textcolor{black}{}} (NominalObserverInput);
			\node [input, right of= StateEstimate1line, node distance=0.8cm] (StateEstimate1) {};
			\node [input, right of= StateEstimate1line, node distance=0.7cm] (StateEstimate1Arrow) {};
			\node [input, left of= StateEstimate1Arrow, node distance=0.54cm] (StateEstimate1ArrowStart) {};
			\draw [draw,->, blue!80](StateEstimate1ArrowStart) -- node  [pos=0.5]{\large \textcolor{black}{$\hat x_1$}} (StateEstimate1Arrow);
			\node [input, below of= NominalObserver, right = 1.7cm, node distance=0.25cm] (MS1lineAbove) {};
			\node [input, below of= MS1lineAbove, node distance=0.25cm] (MS1line) {};
			\node [blockG, right of=MS1line, node distance=3.35cm] (MonitoringSignal1) {
				$\begin{array}{c}
					\textbf{Monitoring} \\ \textbf{variable} \ 1
				\end{array}$
				
			};
			\node [input, above of= MonitoringSignal1, left = 1.49cm, node distance=0.25cm] (MonitoringSignal1Above) {};
			
			\node [input, right of= MS1lineAbove, node distance=0.45cm] (MS1lineAboveArcStart) {};
			\node [input, right of= MS1lineAboveArcStart, node distance=0.4cm] (MS1lineAboveArcEnd) {};
			\draw [draw,-] (MS1lineAbove) -- node {} (MS1lineAboveArcStart);
			\draw [draw,->] (MS1lineAboveArcEnd) -- node  [pos=0.5]{\large $\hat y_1$} (MonitoringSignal1Above);

			\node [input, below of= MonitoringSignal1, left = 1.49cm, node distance=0.25cm] (MonitoringSignal1Below) {};
			\node [input, below of= NominalObserverLine, node distance=0.75cm] (Output1) {};
			
			\node [input, right of= Output1, node distance=4.65cm] (Output1ArcStart) {};
			\node [input, right of= Output1ArcStart, node distance=0.4cm] (Output1ArcEnd) {};
			\draw [draw,-] (Output1) -- node {} (Output1ArcStart);
			\draw [draw,->] (Output1ArcEnd) -- node  [pos=0.5]{\large $ y$} (MonitoringSignal1Below);
			
			\node [input, right of= MonitoringSignal1, node distance=2.5cm] (Eta1) {};
			\node [input, right of= MonitoringSignal1, node distance=2.4cm] (Eta1Arrow) {};
			\draw [draw,->, MyGreen!90] (MonitoringSignal1) -- node  [pos=0.5]{\large \textcolor{black}{$\eta_1$}} (Eta1Arrow);
			
			\node [input, above of= Observer2, right = 1.22cm, node distance=0.25cm] (StateEstimate2line) {};
			\node [input, right of= StateEstimate2line, node distance=1.02cm] (StateEstimate2) {};
			\draw [draw,->, blue!80](StateEstimate2line) -- node  [pos=0.5]{\large \textcolor{black}{$\hat x_2$}} (StateEstimate2);
			\node [input, above of= Observer2, left = 1.23cm, node distance=0.25cm] (Observer2Input) {};
			\node [input, left of= Observer2Input, node distance=1.08cm] (Observer2InputStart) {};
			\node [input, left of= Observer2InputStart, node distance=0.4cm] (Observer2InputArc) {};
			\node [input, left of= Observer2InputArc, node distance=0.5cm] (Observer2InputLeft) {};
			\node [input, above of= Observer2InputLeft, node distance=2cm] (Observer2InputLine) {};
			\draw[black,-] ([yshift=0cm,xshift =-0 cm]Observer2InputArc) arc (180:0:0.2cm);
			\draw [draw,-](Observer2InputArc) -- node {} (Observer2InputLeft);
			\draw [draw,-](Observer2InputLeft) -- node {} (Observer2InputLine);
			
			\draw [draw,->](Observer2InputStart) -- node  [pos=0.5]{\large \textcolor{black}{}} (Observer2Input);
			\node [input, below of= Observer2, right = 1.22cm, node distance=0.25cm] (MS2lineAbove) {};
			\node [input, below of= MS2lineAbove, node distance=0.25cm] (MS2line) {};
			\node [blockG, right of=MS2line, node distance=3.83cm] (MonitoringSignal2) {
				$\begin{array}{c}
					\textbf{Monitoring} \\ \textbf{variable} \ 2
				\end{array}$
			};
			\node [input, above of= MonitoringSignal2, left = 1.49cm, node distance=0.25cm] (MonitoringSignal2Above) {};
			
			\node [input, right of= MS2lineAbove, node distance=0.93cm] (MS2lineAboveArcStart) {};
			\node [input, right of= MS2lineAboveArcStart, node distance=0.4cm] (MS2lineAboveArcEnd) {};
			\draw [draw,-] (MS2lineAbove) -- node {} (MS2lineAboveArcStart);
			\draw [draw,->] (MS2lineAboveArcEnd) -- node  [pos=0.5]{\large $\hat y_2$} (MonitoringSignal2Above);
			
			\node [input, below of= MonitoringSignal2, left = 1.49cm, node distance=0.25cm] (MonitoringSignal2Below) {};
			\node [input, below of= Observer2Line, node distance=0.75cm] (Output2) {};
			
			\node [input, right of= Output2, node distance=4.65cm] (Output2ArcStart) {};
			\node [input, right of= Output2ArcStart, node distance=0.4cm] (Output2ArcEnd) {};
			\draw [draw,-] (Output2) -- node {} (Output2ArcStart);
			\draw [draw,->] (Output2ArcEnd) -- node  [pos=0.5]{\large $ y$} (MonitoringSignal2Below);
			
			\node [input, right of= MonitoringSignal2, node distance=2.4cm] (Eta2) {};
			\draw [draw,->, MyGreen!90] (MonitoringSignal2) -- node  [pos=0.5]{\large \textcolor{black}{$\eta_2$}} (Eta2);
			
			\node [input, above of= ObserverN1, right = 1.22cm, node distance=0.25cm] (StateEstimateN1line) {};
			\node [input, right of= StateEstimateN1line, node distance=1.03cm] (StateEstimateN1) {};
			\draw [draw,->, blue!80](StateEstimateN1line) -- node  [pos=0.5]{\large \textcolor{black}{$\hat x_{N+1}$}} (StateEstimateN1);
			\node [input, above of= ObserverN1, left = 1.23cm, node distance=0.25cm] (ObserverN1Input) {};
			
			\node [input, left of= ObserverN1Input, node distance=1.06cm] (ObserverN1InputStart) {};
			\node [input, left of= ObserverN1InputStart, node distance=0.4cm] (ObserverN1InputArc) {};
			\node [input, left of= ObserverN1InputArc, node distance=2.9cm] (ObserverN1InputLeft) {};
			\draw[black,-] ([yshift=0cm,xshift =-0 cm]ObserverN1InputArc) arc (180:0:0.2cm);
			\draw [draw,-](ObserverN1InputArc) -- node {} (ObserverN1InputLeft);
			\draw [draw,-](inputLine) -- node {} (ObserverN1InputLeft);

			\draw [draw,->](ObserverN1InputStart) -- node  [pos=0.5]{\large \textcolor{black}{}} (ObserverN1Input);
			\node [input, below of= ObserverN1, right = 1.22cm, node distance=0.25cm] (MSN1lineAbove) {};
			\node [input, below of= MSN1lineAbove, node distance=0.25cm] (MSN1line) {};
			\node [blockG, right of=MSN1line, node distance=3.83cm] (MonitoringSignalN1) {
				$\begin{array}{c}
					\textbf{Monitoring} \\ \textbf{variable} \ N+1
				\end{array}$
			};
			\node [input, above of= MonitoringSignalN1, left = 1.49cm, node distance=0.25cm] (MonitoringSignalN1Above) {};
			\draw [draw,->] (MSN1lineAbove) -- node  [pos=0.8]{\large $\hat y_{N+1}$} (MonitoringSignalN1Above);
			\node [input, below of= MonitoringSignalN1, left = 1.49cm, node distance=0.25cm] (MonitoringSignalN1Below) {};
			\node [input, below of= ObserverN1Line, node distance=0.75cm] (OutputN1) {};
			\draw [draw,->] (OutputN1) -- node  [pos=0.9]{\large $y$} (MonitoringSignalN1Below);
			\draw [draw,-] (ObserverN1Line) -- (OutputN1);
			\node [input, above of= StateEstimate1, node distance=0.15cm] (StateEstimate1above) {};
			\node [input, below of= StateEstimateN1, node distance=0.15cm] (StateEstimateN1below) {};
			
			\node [blockMuxBlueBig, below of= StateEstimate1above, node distance=10.3em] (StateEstimateMux) {};
			
			\node [input, right of= MonitoringSignalN1, node distance=2.4cm] (EtaN1) {};
			\draw [draw,->, MyGreen!90] (MonitoringSignalN1) -- node  [pos=0.5]{\large\textcolor{black}{ $\eta_{N+1}$}} (EtaN1);
			\node [input, above of= Eta1, node distance=0.15cm] (Eta1above) {};
			\node [input, below of= EtaN1, node distance=0.15cm] (EtaN1below) {};
			
			\node [blockMuxGreen, below of= Eta1above, node distance=10.3em] (EtaMux) {};
			
			\node at ($(MonitoringSignal2)!.45!(MonitoringSignalN1)$) {\large \textcolor{MyGreen!90}{\vdots}};
			
			\node [input, below of= Eta1, right = 0.12cm, node distance=1cm] (EtaLine) {};
			\node [input, right of= EtaLine, node distance=1.4cm] (Eta) {};
			\draw [draw,-, MyGreen!90] (EtaLine) -- node  [pos=0.6]{\large \textcolor{black}{$\eta$}} (Eta);
			
			\node [blockR, right of=Eta, node distance=2cm] (SigmaBlock) {
				{$	\sigma \in \operatornamewithlimits{\argmin}\limits_{k \in \{1, \dots, N+1\}} \eta_k$}
			};
			\draw [draw,->, MyGreen!90] (Eta) -- (SigmaBlock);
			\node [input, above of= SigmaBlock, right = 1.55cm, node distance=0.7cm] (SigmaBlockName) {};
			\draw [] (SigmaBlockName) -- node  [pos=0.5]{\textbf{Selection Criterion}} (SigmaBlockName);
			
			\node [input, below of= StateEstimate1, right = 0.12cm, node distance=4cm] (StateEstimateLine) {};
			\node [input, right of= StateEstimateLine, node distance=6.3cm] (StateEstimate) {};
			
			\node [input, right of= StateEstimateLine, node distance=4.88cm] (StateEstimateArcStart) {};
			\node [input, right of= StateEstimateArcStart, node distance=0.4cm] (StateEstimateArcEnd) {};
			\draw[blue!80,-] ([yshift=0cm,xshift =-0 cm]StateEstimateArcStart) arc (180:0:0.2cm);
			\draw [draw,-, blue!80] (StateEstimateLine) -- node {} (StateEstimateArcStart);
			\node [input, left of= StateEstimate, node distance=0.1cm] (StateEstimateArrow) {};
			\draw [draw,->, blue!80] (StateEstimateArcEnd) -- node  [pos=0.5]{\large \textcolor{black}{$\hat x$}} (StateEstimateArrow);
			
			\node [input, above of= StateEstimate, node distance=1cm] (StateEstimate1bisLine) {};
			\node [input, above of= StateEstimate, node distance=0.3cm] (StateEstimate2bisLine) {};
			\node [input, below of= StateEstimate, node distance=1cm] (StateEstimateN1bisLine) {};

			\node [blockMuxBlueSmall, below of= StateEstimate1bisLine, node distance=2.85em] (StateEstimateBisMux) {};
			
			\node [input, right of= StateEstimate1bisLine, node distance=1cm] (StateEstimate1bis) {};
			\node [input, right of= StateEstimate2bisLine, node distance=1cm] (StateEstimate2bis) {};
			\node [input, right of= StateEstimateN1bisLine, node distance=1cm] (StateEstimateN1bis) {};
			\node [input, right of= StateEstimate1bisLine, node distance=0.12cm] (StateEstimate1bisArrow) {};
			\node [input, right of= StateEstimate2bisLine, node distance=0.12cm] (StateEstimate2bisArrow) {};
			\node [input, right of= StateEstimateN1bisLine, node distance=0.12cm] (StateEstimateN1bisArrow) {};
			\draw [draw,-, blue!80] (StateEstimate1bisArrow) -- node  [pos=0.5]{\large \textcolor{black}{$\hat x_1$}} (StateEstimate1bis);
			\draw [draw,-, blue!80] (StateEstimate2bisArrow) -- node  [pos=0.5]{\large \textcolor{black}{$\hat x_2$}} (StateEstimate2bis);
			\draw [draw,-, blue!80] (StateEstimateN1bisArrow) -- node  [pos=0.6]{\large \textcolor{black}{$\hat x_{N+1}$}} (StateEstimateN1bis);
			\draw [fill=black] (StateEstimate1bis) circle (2pt);
			\draw [fill=black] (StateEstimate2bis) circle (2pt);
			\draw [fill=black] (StateEstimateN1bis) circle (2pt);
			
			\node at ($(StateEstimate2bis)!.4!(StateEstimateN1bis)$) {\large \vdots};
			
			\node [input, right of= StateEstimate, node distance=2.5cm] (Switching) {};
			\draw [fill=black] (Switching) circle (2pt);
			\draw [draw,-] (StateEstimate1bis) -- (Switching);
			
			\node [input, below of= SigmaBlock, node distance=1.7cm] (SigmaSwitch){};
			\draw [draw,->] (SigmaBlock) -- node  [pos=0.5]{\large $\sigma$} (SigmaSwitch);
			
			\draw[black,-, densely dotted, line width=0.25mm] ([yshift=.5cm,xshift = -0.4cm]Switching) arc (115:245:0.45cm);
			\draw [black,-latex] (17.10,-0.23)to (17.20,-0.17) ;
			\draw [black,-latex] (17.10,-1.08)to (17.20,-1.14) ;
			
			\node [input, right of= SigmaBlock, node distance=2.8cm] (EtaSigma){};
			\draw [draw,->, MyGreen!90] (SigmaBlock) -- node  [pos=0.5]{\large \textcolor{black}{$\eta_\sigma$}} (EtaSigma);
			
			\node [input, right of= Switching, node distance=2.6cm] (hatXSigma){};
			\draw [draw,->, blue!80] (Switching) -- node  [pos=0.5]{\large \textcolor{black}{$	\hat x_\sigma$}} (hatXSigma);
			
			\node [input, left of= EtaSigma, node distance=0.4cm] (EtaSigmaFeedback){};
			\node [input, below of= EtaSigmaFeedback, node distance=5cm] (EtaSigmaFeedbackBelow1){};
			\node [input, below of= EtaSigmaFeedback, node distance=7.5cm] (EtaSigmaFeedbackBelow2){};
			\draw [draw,-, densely dotted, line width=0.2mm, MyGreen!90] (EtaSigmaFeedback) -- (EtaSigmaFeedbackBelow2);
			
			\node [input, below of= MonitoringSignal2, right = 0.3cm, node distance=4cm] (EtaSigmaFeedbackMS2){};
			\draw [draw,-, densely dotted, line width=0.2mm, MyGreen!90] (EtaSigmaFeedbackBelow1) -- (EtaSigmaFeedbackMS2);
			\node [input, right of= MonitoringSignal2, below = 0.63cm, node distance=0.3cm] (MonitoringSignal2right){};
			\draw [draw,->, densely dotted, line width=0.2mm, MyGreen!90] (EtaSigmaFeedbackMS2) -- node  [pos=0.9]{\large \textcolor{black}{$\eta_\sigma$}} (MonitoringSignal2right);
			
			\node [input, below of= MonitoringSignalN1, right = 0.3cm, node distance=1.5cm] (EtaSigmaFeedbackMSN1){};
			\draw [draw,-, densely dotted, line width=0.2mm, MyGreen!90] (EtaSigmaFeedbackBelow2) -- (EtaSigmaFeedbackMSN1);
			\node [input, right of= MonitoringSignalN1, below = 0.63cm, node distance=0.3cm] (MonitoringSignalN1right){};
			\draw [draw,->, densely dotted, line width=0.2mm, MyGreen!90] (EtaSigmaFeedbackMSN1) -- node  [pos=0.3]{\large \textcolor{black}{$\eta_\sigma$}} (MonitoringSignalN1right);
			
			\node [input, left of= hatXSigma, node distance=0.7cm] (hatXSigmaFeedback){};
			\node [input, below of= hatXSigmaFeedback, node distance=2cm] (hatXSigmaFeedbackBelow1){};
			\node [input, below of= hatXSigmaFeedback, node distance=4.8cm] (hatXSigmaFeedbackBelow2){};
			\draw [draw,-, densely dotted, line width=0.2mm, blue!80] (hatXSigmaFeedback) -- (hatXSigmaFeedbackBelow2);
			
			\node [input, below of= Observer2, right = 0.3cm, node distance=3.8cm] (hatXSigmaFeedback2){};
			\draw [draw,-, densely dotted, line width=0.2mm, blue!80] (hatXSigmaFeedbackBelow1) -- (hatXSigmaFeedback2);
			\node [input, right of= Observer2, below = 0.63cm, node distance=0.3cm] (Observer2right){};
			\draw [draw,->, densely dotted, line width=0.2mm, blue!80] (hatXSigmaFeedback2) -- node  [pos=0.8]{\large \textcolor{black}{$\hat x_\sigma$}} (Observer2right);
			
			\node [input, below of= ObserverN1, right = 0.3cm, node distance=1.5cm] (hatXSigmaFeedbackN1){};
			\draw [draw,-, densely dotted, line width=0.2mm, blue!80] (hatXSigmaFeedbackBelow2) -- (hatXSigmaFeedbackN1);
			\node [input, right of= ObserverN1, below = 0.63cm, node distance=0.3cm] (ObserverN1right){};
			\draw [draw,->, densely dotted, line width=0.2mm, blue!80] (hatXSigmaFeedbackN1) -- node  [pos=0.3]{\large \textcolor{black}{$\hat x_\sigma$}} (ObserverN1right);
			
			\draw[black,-] ([yshift=0cm,xshift =-0 cm]MS1lineAboveArcStart) arc (180:0:0.2cm);
			\draw[black,-] ([yshift=0cm,xshift =-0 cm]Output1ArcStart) arc (180:0:0.2cm);
			\draw[black,-] ([yshift=0cm,xshift =-0 cm]MS2lineAboveArcStart) arc (180:0:0.2cm);
			\draw[black,-] ([yshift=0cm,xshift =-0 cm]Output2ArcStart) arc (180:0:0.2cm);
			
		\end{tikzpicture}
	\end{center}
	\caption{Block diagram representing the system architecture. $\eta:= (\eta_1, \dots \eta_{N+1})$, $\hat x:= (\hat x_1, \dots, \hat x_{N+1})$.}
	\label{Fig:blockDiagram}
\end{figure*}
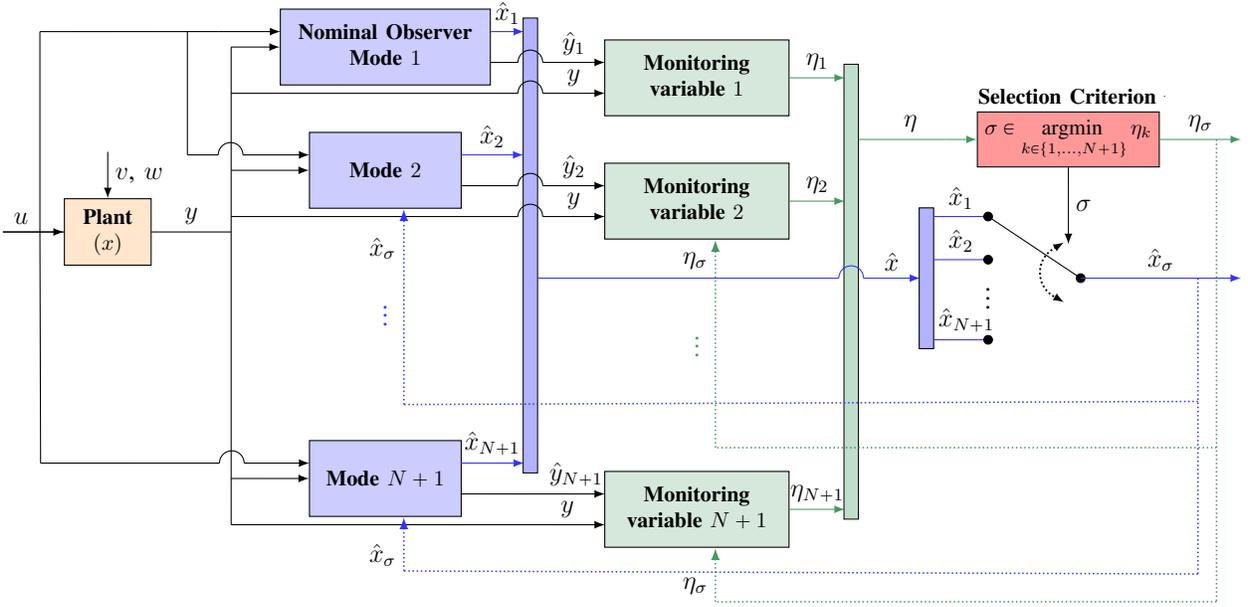

The hybrid estimation scheme we propose consists of the following elements, see Fig.~\ref{Fig:blockDiagram}:
\begin{itemize}
	\item \textit{nominal observer} given in \eqref{eq:observerNominal}; 
	\item $N$ additional dynamical systems of the form of \eqref{eq:observerNominal} but with a different output injection gain. 
	Each of these systems, as well as the nominal observer, is called \textit{mode} for the sake of convenience;
	\item \textit{monitoring variables} used to evaluate the performance of each mode of the multi-observer;
	\item \textit{selection criterion}, that switches between the state estimates produced by the different modes exploiting the performance knowledge given by the monitoring variables;
	\item \textit{reset rule}, that explains how the estimation scheme may be updated when we switch the selected mode. 
\end{itemize}
All these elements together form the hybrid multi-observer. We describe each component in the sequel.

\subsection{Additional modes} \label{ObserverLikeSystemsSubsection} 
We consider $N$ additional dynamical systems, where the integer $N \in \Zp$ is arbitrarily selected by the user. 
These $N$ extra systems are of the form of \eqref{eq:observerNominal} but with a different output injection gain, i.e., for any $k\in \{2, \dots, N+1\}$, the $k^{\text{th}}$ mode of the multi-observer is given by
\begin{equation}
	\begin{aligned}
		\dot{\hat{x}}_k  &= f_o(\hat{x}_k,u, L_k (y-\hat{y}_k))\\
		\hat{y}_k  &=  h(\hat{x}_k, 0),
	\end{aligned}
	\label{eq:observer}
\end{equation}
where $\hat{x}_k \in \R^{n_x}$ is the $k^{\text{th}}$ mode state estimate, $\hat{y}_k \in \R^{n_y}$ is the $k^{\text{th}}$ mode output and $L_k \in \R^{n_{L_1} \times{n_y}}$ is its gain. 
 It is important to emphasize that we make no assumptions on the convergence properties of the solution to system \eqref{eq:observer} contrary to observer \eqref{eq:observerNominal}. Thus  we have full freedom for selecting the gains $L_k \in \R^{{n_{L_1}}\times n_y}$, with $k \in \{2, \dots, N+1\}$. 
We will elaborate more on the choice of the $L_k$'s in Section~\ref{DesignGuidelinesSubsection}.
\begin{rem}
	We also have full freedom in the choice of the initial conditions of all modes of the multi-observer. 
	In practice, we can select all of them equal, but it is not necessary for the stability results in Section~\ref{Main result}  to hold. 
	\hfill$\Box$
\end{rem}

\subsection{Monitoring variables}\label{MonitoringVariables}

The idea is to select the ``best" mode among the $N+1$ of the multi-observer at all time instants, in a sense that will be made precise below, with the aim of  improving the estimation performance provided by observer~\eqref{eq:observerNominal}. Ideally, the criterion used to evaluate the performance of each mode would depend on the estimation errors  $e_k = x- \hat{x}_k$, with $k \in \{1, \dots, N+1\}$. 
However, since the state $x$ is unknown, $e_k$ is unknown and any performance criterion involving $e_k$ would not be implementable. Instead, as done in other contexts, see e.g.,  \cite{willems2004deterministic, na2017adaptive}, we rely on the knowledge of the output $y$ and the estimated outputs $\hat{y}_k$ for $k\in \{1,\dots, N+1\}$. 
In particular, inspired by \cite{willems2004deterministic}, to evaluate the performance of each mode,  we 
introduce the monitoring variable $\eta_k\in \R_{\geq 0}$ for any $k\in\{1, \dots, N+1\}$, whose dynamics is 
%
\begin{equation}
	\begin{aligned}
	\dot{\eta}_k &= -\nu\eta_k + \lambda_1|y-\hat{y}_k|^2 + \lambda_2|L_k(y-\hat{y}_k)|^2\\ 
	&=: g(\eta_k, L_k, y, y_k ), 
	\end{aligned}
	\label{eq:etaDynamics}
\end{equation}
where $\lambda_1, \lambda_2 \in \R_{\geq 0}$, with $\max(\lambda_1,\lambda_2)>0$ and $\nu \in(0,\alpha]$ are design parameters, and $\alpha$ comes from Assumption~\ref{NominalAssumption}. 
The term $\lambda_1 |y-\hat y_k|^2$ in \eqref{eq:etaDynamics} is related to the output estimation error, while $\lambda_2 |L_k(y-\hat y_k)|^2$ takes into account the correction effort of the observer, also called latency in \cite{willems2004deterministic}.
Note that the monitoring variable $\eta_k$ in \eqref{eq:etaDynamics} for all $k \in \{1 \dots, N+1\}$ is implementable since we have access to the output $y$ and all the estimated outputs $\hat{y}_k$ at all time instants. 
\subsection{Selection criterion}\label{SelectionCriterion}
Based on the monitoring variables $\eta_k$, with $k \in \{1, \dots, N+1\}$, which provide evaluations of the performance of all the modes of the multi-observer, we define a criterion to select the state estimate to look at. We use a signal $\sigma:\R_{\geq0} \to \{1, \dots, N+1\}$ for this purpose, and we denote the selected state estimate mode $\hat x_\sigma$ and the associated monitoring variable $\eta_\sigma$. 
To improve the estimation performance of observer \eqref{eq:observerNominal} we should select the mode that minimizes its monitoring variable $\eta_k$, with $k \in \{1, \dots, N+1\}$. Indeed, equation \eqref{eq:etaDynamics} implies   that for any $k \in \{1, \dots, N+1\}$, for any initial condition $\eta_k (0) \in \R_{\geq 0}$, 
for any $y, \hat{y}_k \in \mathcal{L}_{\R^{n_y}}$, and any $t\geq 0$, 
\begin{equation}
	\begin{aligned}
		\eta_k(t) = & \ e^{-\nu t}\eta_k(0) + 
		\int_{0}^{t}e^{-\nu(t-\tau)}\left( \lambda_1|y(\tau)-\hat{y}_k(\tau)|^2 \right. \\
		&+  \left. \lambda_2|L_k(y(\tau)-\hat{y}_k(\tau))|^2  \right)d\tau.
	\end{aligned}
	\label{eq:etaDynamicsTime}
\end{equation}
Consequently, selecting the mode with the minimal monitoring variable implies minimizing the cost \eqref{eq:etaDynamicsTime} over the modes $k \in \{1, \dots, N+1\}$. 
 However, this selection criterion may produce many switching. 
 To mitigate this phenomenon\footnote{The hybrid model presented in Section \ref{HybridModel} may generate Zeno solutions. We argue that this is not an issue as we can also prove that complete solutions with unbounded continuous-time domains exists for any initial conditions, under mild assumptions, so that Zeno can always be avoided in practice. This will be detailed in a future work.}, we introduce a parameter $\varepsilon \in (0,1]$ and we propose to switch the selected mode only when there exists $k \in \{1,\dots, N+1\} \setminus \{\sigma\}$ such that $\eta_k \leq \varepsilon \eta_{\sigma}$. Hence, when $\varepsilon<1$, the idea is to wait that the current minimum $\eta_k$ is significantly smaller than $\eta_\sigma$ before updating $\sigma$. 
%
In that way, at the initial time $t_0 = 0$,  we take $\sigma(0) \in \operatornamewithlimits{\argmin}\limits_{k \in \{1, \dots, N+1\}} \eta_k(0)$. Then, $\sigma$ is kept constant, i.e., $\dot \sigma (t) =0$ for all $t \in(0, t_{1})$, with $t_{1}:= \inf \{t \geq 0: \exists k \in \{1,\dots, N+1\} \setminus \{\sigma(t)\}  \text{ such that } \eta_k(t) \leq \varepsilon \eta_{\sigma(t)}(t)\}$. At time $t_1$ we switch the selected mode according to $\displaystyle \sigma(t_{1}) \in \operatornamewithlimits{\argmin}\limits_{k \in \{1, \dots, N+1\}} \eta_k(t_{1})$.
We repeat these steps iteratively and we denote with $t_i \in \R_{\geq 0}$, $i \in \Zp$ the $i^{\textnormal{th}}$ time when the selected mode changes (if it exists), i.e.,  $t_{i}:= \inf \{t \geq t_{i-1}: \exists k \in \{1,\dots, N+1\} \setminus \{\sigma(t)\}  \text{ such that } \eta_k(t) \leq \varepsilon \eta_{\sigma(t)}(t)\}$. Consequently, for all $i\in \Zo$, $\dot \sigma(t) = 0$ for all $t \in (t_{i-1}, t_{i})$ and $\displaystyle \sigma(t_{i}) \in \operatornamewithlimits{\argmin}\limits_{k \in \{1, \dots, N+1\}} \eta_k(t_{i})$.
It is important to notice that, if the monitoring variables of more than one mode have the same value and it is the minimum between all the $\eta_k$, with $k \in \{1, \dots, N+1\}$, the proposed technique selects randomly one of them and this is not an issue for the stability results presented later in Section~\ref{Main result}. 
\color{black}
\begin{rem}
	The scheme proposed in this paper works for any initial condition  $\eta_k(0) \in \R_{\geq 0}$, 
	for all $k \in \{1, \dots, N+1\}$, which corresponds to the initial cost of each mode of the multi-observer. 
	Consequently, the choice of $\eta_k(0)$ is an extra degree of freedom that can be used to initially penalize the modes when there is a prior knowledge of which mode should be initially selected. Conversely, in the case where there is not prior knowledge of which mode should be chosen at the beginning, all $\eta_k$, with $k \in \{1, \dots, N+1\}$, can be initialized at the same value such that the term $e^{-\nu t}\eta_k(0)$ in \eqref{eq:etaDynamicsTime} is irrelevant for the minimization.
	\hfill$\Box$
\end{rem}
\color{black}

\subsection{Reset rule}\label{ResetRule}
When a switching occurs, i.e., when a different mode is selected, 
we propose two different options to update the hybrid estimation scheme. The first one, called \textit{without resets}, consists in only updating $\sigma$, and consequently, we only switch the state estimate we are looking at. Conversely, the second option, called \textit{with resets}, consists in not only switching the mode that is considered, but also resetting the state estimates and the monitoring variables of all the modes $k \in \{2, \dots, N+1\}$ to the updated $\hat x_\sigma$ and $\eta_\sigma$ respectively. 
Only the state estimate and the monitoring variable of the nominal observer~\eqref{eq:observerNominal}, corresponding to mode $1$, are not reset. 

We use the parameter $r \in \{0,1\}$ to determine which option is selected, where $r = 0$ corresponds to the case without resets, while $r = 1$ corresponds to the case where the resets are implemented. The state estimate $\hat x_k$ of the $k^{\text{th}}$ mode, with $k \in \{2, \dots, N+1\}$ and its monitoring variable $\eta_k$, when a switch of the considered mode occurs, are thus defined as, at a switching time $t_i \in \R_{\geq 0}$, 
\begin{equation}
	\begin{aligned}
		&\hat x_k(t_i^+) \in \hat \ell_k (\hat x(t_i), \eta(t_i)) \\
		&\quad := \{(1-r)\hat x_k(t_i) + r \hat x_{k^\star}(t_i): 
		k^\star \in  \operatornamewithlimits{\argmin}\limits_{j \in \{ 1, \dots, N +1\}} \eta_j(t_i) \},
	\end{aligned}
	\label{eq:StateEstimatek_plus}
\end{equation}
\begin{equation}
	\begin{aligned}
		&\eta_k(t_i^+) \in p_k (\eta(t_i)) \\
		& \quad := \{(1-r)\eta_k(t_i) + r \eta_{k^\star}(t_i): 
		k^\star \in  \operatornamewithlimits{\argmin}\limits_{j \in \{ 1, \dots, N +1\}} \eta_j(t_i) \},
	\end{aligned}
	\label{eq:Eta_plus}
\end{equation}
where $\hat x:= (\hat x_1, \dots, \hat x_{N+1})$ and $\eta:= (\eta_1, \dots, \eta_{N+1})$.
Note that, if the monitoring variables of more than one mode have the same value and it is the minimum between all the $\eta_k$, with $k \in \{1, \dots, N+1\}$, then, from \eqref{eq:StateEstimatek_plus}, the modes may be reset with different state estimates. 
%
%

	Note that, when $\varepsilon = 1$, with the proposed technique we have $\eta_{\sigma(t)}(t) \leq \eta_1(t)$ for all $t \geq 0$, both in the case without and with resets. Therefore, the estimation performance of the proposed hybrid multi-observer are always not worse than the performance of the nominal one, according to the cost that we consider. 

\subsection{Design guidelines}\label{DesignGuidelinesSubsection}
We summarize the procedure to follow to design the hybrid estimation scheme. 
\begin{enumerate}
		\item  Design the nominal observer \eqref{eq:observerNominal} such that Assumption~\ref{NominalAssumption} holds. 
		\item  Select $N$ gains $L_2, \dots, L_{N+1}$ 
			for the $N$ additional modes in  \eqref{eq:observer}.
		\item  Implement in parallel the $N+1$ modes of the multi-observer.
		\item Generate the monitoring variables $\eta_k$, with $k \in \{1, \dots, N+1\}$. 
		\item Select $\varepsilon \in (0, 1]$ and evaluate the signal $\sigma$.
		\item  Run the hybrid scheme without or with resets.
		\item  $\hat x_\sigma$ is the state estimate to be considered for estimation purpose. 
	\label{DesignStepsAlgorithm}
\end{enumerate}

There is a lot of flexibility in the number of additional modes $N$ and the selection of the gains $L_k$, with $k \in \{2, \dots, N+1\}$. This allows to address the different trade-offs of the state estimation of nonlinear systems. 
We believe that the gains selection
 has to be done on a case-by-case basis since it is related to the structure of the nominal observer and it depends on the available computational capabilities. For example, when the nominal observer \eqref{eq:observerNominal} is a high-gain observer,  see e.g., \cite{khalil2014high, astolfi2015high} or, more generally, an infinite gain margin observer \cite[Section 3.4]{bernard2022observer}, we typically need to select a very large gain based on a conservative bound to ensure Assumption~\ref{NominalAssumption}, which would result in 
fast convergence of the estimation error, but, unfortunately, it will be very sensitive to measurement noise. In this case, to overcome the conservatism of the theory, an option is to select the $L_k$ gains (much) smaller than the nominal one, even though there is no convergence proof for these choices, in order to obtain a state estimate which is more robust to measurement noise. This is the approach followed in Section~\ref{Example}. 
In general, possible options to select the additional gains are to pick them in a neighborhood of the nominal one, or to scale the nominal gain by some factors. This gain selection will produce systems with different behaviors and switching between them should allow an improvement of the estimation performance, as illustrated on a numerical example in Section~\ref{Example}. A careful analysis of how these gains should be selected is left for future work.
%
\section{Hybrid model}\label{HybridModel}
To proceed with the analysis of the hybrid estimation scheme presented in Section IV, we model the overall system as a hybrid system of the form of \cite{goebel2012hybrid}, where a jump corresponds to a switch of the selected mode and a possible reset as explained in Section~\ref{ResetRule}. We define the overall hybrid state as $q:= (x, \hat{x}_1, \dots, \hat{x}_{N+1},\eta_1, \dots, \eta_{N+1}, \sigma) \in \mathcal{Q}:= \R^{n_x} \times \R^{(N+1)n_x} \times \R_{\geq 0}^{N+1}  \times \{1, \dots, N +1\}$,
and we obtain the hybrid system 
\begin{equation}
	\left\lbrace \
	\begin{aligned}
		\dot{q} &= F(q, u, v, w), \ \ \ \ \ &&q \in \mathcal{C}\\
		q^{+} &\in G(q), \ \ \ \ \ &&q \in \mathcal{D},
	\end{aligned}
	\right.
	\label{eq:HybridSystemGeneral}
\end{equation}
where flow map is defined as, for any $q \in \mathcal{C}$, $u \in \mathcal{U}$, $v \in \mathcal{V}$ and $w \in \mathcal{W}$, from \eqref{eq:system}, \eqref{eq:observerNominal}, \eqref{eq:observer}, \eqref{eq:etaDynamics}, $	F(q, u,v,w):=
(f_p(x,u, v), f_o(\hat x_1,u,L_1(y-\hat{y}_1)), \dots, f_o(\hat x_{N+1},u,L_{N+1}(y-\hat{y}_{N+1})), g(\eta_1, L_1, y, y_1 ), $ $ \dots, g(\eta_{N+1}, L_{N+1}, y, y_{N+1}), 0)$.
%
%
%
The jump map in \eqref{eq:HybridSystemGeneral} is defined as, for any $q \in \mathcal{D}$, \textcolor{black}{ from \eqref{eq:StateEstimatek_plus}, \eqref{eq:Eta_plus},} 
$	G(q) :=( 
x,
\hat x_1,
\hat \ell_2(q),
\dots, 
\hat \ell_{N+1} (q),
\eta_1,
p_2(q),
\dots,
p_{N+1} (q),\\
\operatornamewithlimits{\argmin}\limits_{k \in \{ 1, \dots, N +1\}} \eta_k)$.
%
%
%
In view of Section \ref{SelectionCriterion}, the flow and jump sets $\mathcal{C}$ and $\mathcal{D}$ in \eqref{eq:HybridSystemGeneral} are defined as 
\begin{align}
	\mathcal{C} := \left\{q \in \mathcal{Q}: \forall k \in \{1, \dots, N+1\} \:  \ \eta_k \geq  \varepsilon\eta_\sigma \right\}\!,	\label{eq:flowSet} \\
	\mathcal{D} := \left\{q \in \mathcal{Q}: \exists k \in \{1, \dots, N+1\} \setminus \{\sigma\} \: \ \eta_k \leq \varepsilon\eta_\sigma \right\}\!.
	\label{eq:jumpSet}
\end{align}

\section{Stability guarantees}\label{Main result}
The goal of this section is to prove that the proposed hybrid estimation scheme satisfies an input-to-state stability property. 
For this purpose, we first make an assumption on the output map. 

\subsection{Assumption on the output map}\label{ass1Subsection}
We make the next assumption on the output map $h$ in \eqref{eq:system}.
\begin{ass}
	There exist $\delta_1, \delta_2 \in \R_{> 0}$ such that for all $x, x' \in \R^{n_x}$, $w, w' \in \mathcal{W}$,  
	\begin{equation}
		|h(x,w) - h(x', w')|^2 \leq \delta_1 V(x-x') + \delta_2|w-w'|^2,
		\label{eq:equationAssumptionOutputMap}
	\end{equation}
	where $V$ comes from Assumption~\ref{NominalAssumption}. \hfill $\Box$
	\label{ASS:ass1}
\end{ass}
Assumption~\ref{ASS:ass1} holds in the common case where $V$ in Assumption~\ref{NominalAssumption} is quadratic and $h$ is globally Lipschitz. Indeed, in this case, 
$V(x-x'):= (x-x')^\top P(x-x')$, with $P \in \R^{n_x \times n_x}$ symmetric positive definite, and 
$|h(x,w) - h(x',w')| \leq K|(x-x', w-w')|$ for any $x, x' \in \R^{n_x}$, $w, w' \in \mathcal{W}$ and some $K \in \R_{\geq 0}$, then \eqref{eq:equationAssumptionOutputMap} holds with $\displaystyle \delta_1 = \frac{K^2}{\lambda_{\min}(P)}$ and $\displaystyle \delta_2 = K^2$. Note that $h$ globally Lispchitz covers the common case where  $h(x,w)=Cx+Dw$ with $C\in\R^{n_y \times n_x}$ and $D\in \R^{n_y \times n_w}$.


\subsection{Input-to-state stability}\label{StabilitySubsectionSolutions}
In the next theorem we prove that system \eqref{eq:HybridSystemGeneral}-\eqref{eq:jumpSet} satisfies an input-to-state stability property. 
\begin{thm}
	Consider system \eqref{eq:HybridSystemGeneral}-\eqref{eq:jumpSet} and suppose Assumptions~\ref{NominalAssumption}-\ref{ASS:ass1} hold. 
	Then there exist $\beta_U \in \KL$ and $\gamma_U \in \Kinf$ such that for any input $u \in \mathcal{L_U}$, disturbance input $v \in \mathcal{L_V}$ and measurement noise $w \in \mathcal{L_W}$, any solution $q$ satisfies 
	\begin{equation}
		\begin{array}{l}
			|(e_1(t,j), \eta_1(t,j), e_{\sigma} (t,j), \eta_{\sigma}(t,j))|  \\[0.5em]
			\qquad \leq \beta_U(|(e(0,0), \eta(0,0))|,t) + \gamma_U(\norm{v}_{[0,t]} + \norm{w}_{[0,t]})
		\end{array}
	\label{eq:LyapunovSolutionTheoremEquation}
	\end{equation}
	for all $(t,j) \in \dom q$, with $e:= (e_1, \dots, e_{N+1})$ and $\eta:= (\eta_1, \dots, \eta_{N+1})$.
	\hfill $\Box$
	\label{Prop:LyapunovSolutionProposition}
\end{thm}
%
Theorem~\ref{Prop:LyapunovSolutionProposition} guarantees a two-measure input-to-state stability property \cite{cai2007smooth}. 
In particular, \eqref{eq:LyapunovSolutionTheoremEquation} ensures that $e_1$, $\eta_1$,  $e_\sigma$ and $\eta_\sigma$ converge to a neighborhood of the origin, whose ``size'' depend on the $\mathcal{L}_\infty$ norm of $v$ and $w$. Note that we do not guarantee any stability property for the modes $k \neq \sigma$, but this is not needed for the convergence of the hybrid observer estimation error $e_\sigma$. 
Hence, the convergence of the estimated state vector of the selected mode is guaranteed by Theorem~\ref{Prop:LyapunovSolutionProposition}, and recall that, in terms of performance, we have the guarantee by design that $\eta_\sigma \leq \eta_1$ along any solutions to system \eqref{eq:HybridSystemGeneral}-\eqref{eq:jumpSet} when $\varepsilon = 1$, see Section \ref{ResetRule}. 
  
The proof of Theorem~\ref{Prop:LyapunovSolutionProposition} is given in the appendix and relies on the Lyapunov properties established in Section~\ref{LyapunovSubsection}.

\subsection{Lyapunov properties}\label{LyapunovSubsection}
In this section we state the Lyapunov properties,  which are exploited to prove Theorem~\ref{Prop:LyapunovSolutionProposition} in the Appendix.
%
Based on  Assumption~\ref{NominalAssumption}, we prove the next input/output-to-state stability property \cite{sontag2008input} for the estimation error system $e := x- \hat{x} \in \R^{n_x}$ associated to \eqref{eq:system} and \eqref{eq:observer}, whose dynamics is defined as 
\begin{equation}
	\begin{aligned}
	\dot e &= f_p(x,u,v)- f_o(\hat{x},u,  L(y - \hat{y}))\\
	&:= \bar f(e, x,u, v, w, L).
	\end{aligned}
\label{eq:estimationErrorGeneralForIOSS}
\end{equation}
\begin{lem}\label{PropositionOSSobservers}
	Suppose Assumption~\ref{NominalAssumption} holds. 
	Then, for any $x \in \R^{n_x}$, $u \in  \mathcal{U}$, $v \in  \mathcal{V}$, $w \in  \mathcal{W}$, $\hat{x} \in \R^{n_x}$ and any $L \in \R^{{n_{L_1}} \times n_y}$, 
	\begin{equation}
	\begin{array}{l}
		\left\langle \nabla V(e),\bar f(e, x,u, v, w, L)\big) \right\rangle \\
		\quad \leq -\alpha V(e) + \psi_1(|v|) + \psi_2(|w|) + \gamma\norm{L-L_1}^2 |y- \hat{y}|^2,\\
	\end{array}
	\label{eq:PropositionOSS}
\end{equation}
	with 
	$\hat{y} = h(\hat{x}, 0) \in \R^{n_{y}}$ and $\alpha, \psi_1, \psi_2, \gamma, V$ 
	come from Assumption~\ref{NominalAssumption}. 
	\hfill $\Box$
\end{lem}
\color{red}
\color{black}

Lemma~\ref{PropositionOSSobservers} implies that, for $e_k:=x-\hat x_k$ for any $k\in \{2, \dots, N+1\}$, the $e_k$-system, which follows from \eqref{eq:system} and \eqref{eq:observer}, satisfies an input/output-to-state property \cite{sontag2008input} with the same Lyapunov function as in Assumption~\ref{NominalAssumption} for any choice for the observer gain $L_k \in \R^{{n_{L_1}} \times n_y}$. 
The major difference between \eqref{eq:NominalAssumptionDerivative} and \eqref{eq:PropositionOSS} is the term $ \gamma||(L - L_1)||^2|y- \hat{y}|^2$ in \eqref{eq:PropositionOSS}, which may have a destabilizing effect and thus may prevent the $e_k$-system to exhibit input-to-state stability properties similar to \eqref{eq:ISSnominalConverging}. 
	In the next proposition we prove a Lyapunov stability property for system \eqref{eq:HybridSystemGeneral}-\eqref{eq:jumpSet}, whose proof is omitted for space reasons. 

\begin{prop}
	Suppose Assumptions~\ref{NominalAssumption}-\ref{ASS:ass1} hold.  
	Given any sets of gains  $L_k \in \R^{{n_{L_1}} \times{n_y}}$, with $k \in \{2, \dots, N+1\}$, $\nu \in (0,\alpha]$ and $\lambda_1, \lambda_2 \in \R_{\geq 0}$, with $\max(\lambda_1,\lambda_2)>0$, there exist $U:\mathcal{Q} \to \R_{\geq 0}$ locally Lipschitz, and 
	$\underline{\alpha}_U, \overline{\alpha}_U \in \Kinf$, $\alpha_0 \in \R_{> 0}$, $\phi_1, \phi_2 \in \Kinf$, such that
	the following properties hold. 
	\begin{enumerate}[label=(\roman*),leftmargin=.6cm]
		\item For any $q \in \mathcal{Q}$, 
	$\underline{\alpha}_U(|(e_1, \eta_1, e_\sigma, \eta_\sigma)|) \leq U(q) \leq \overline{\alpha}_U(|( e, \eta)|)$,
with $e:= (e_1, \dots, e_{N+1})$ and $\eta:= (\eta_1, \dots, \eta_{N+1})$. 
\item For any $q \in \mathcal{C}$, $u \in \mathcal{U}$, $v \in \mathcal{V}$ and $w \in \mathcal{W}$, such that $F(q,u,v,w) \in T_\mathcal{C}(q)$,
	$U^\circ(q,F(q,u,v,w)) \leq -\alpha_0 U(q) + \phi_1(|v|) + \phi_2(|w|)$.
\item For any $q \in \mathcal{D}$, for any $\mathfrak{g} \in G(q)$, 
	$U(\mathfrak{g} ) \leq U(q)$. \hfill $\Box$
	\end{enumerate}
	\label{THM:LyapunovTheorem}
\end{prop}

The Lyapunov Function $U$ used in the proof is given by $U(q) := c_1 (aV(e_1) + \eta_1) + c_2 \max\limits_{k \in \{1, \dots, N+1\}}\{bV(e_k) - \eta_k, 0\} + c_3 \max\{\varepsilon \eta_\sigma - \eta_1, 0\}$ for any $q \in \mathcal{Q}$, with suitably selected parameters $c_1, c_2, c_3, a, b \in \R_{> 0}$. 
%
Proposition~\ref{THM:LyapunovTheorem} shows the existence of a Lyapunov function $U$ for system \eqref{eq:HybridSystemGeneral}-\eqref{eq:jumpSet}, which is used to prove the input-to-state stability property in Theorem~\ref{Prop:LyapunovSolutionProposition}. 

\section{Numerical case study}\label{Example}
In this section we design the hybrid estimation scheme for the estimation of a Van der Pol oscillator 
\begin{equation}
	\begin{aligned}
		\dot{x} &= Ax + B\varphi(x)\\
		y &= Cx + w
	\end{aligned}
	\label{eq:VanDerPolSystem}
\end{equation}
where $x =(x_1, x_2) \in \R^2$ is the system state to be estimated, $y \in \R$ is the measured output and $w \in \R$ is the measurement noise. The system matrices are
\begin{equation}
	\begin{array}{l}
		A = \begin{bmatrix}
			0 & 1 \\
			0 & 0 \\
		\end{bmatrix}, \
		B = \begin{bmatrix}
			0 \\
			1  \\
		\end{bmatrix}, \
		C = \begin{bmatrix}
			1 & 0
		\end{bmatrix}
	\end{array}
	\label{eq:VanDerMatrices}
\end{equation}
and $\varphi(x) = \sat( - x_1 + 0.5(1-x_1^2)x_2)$ for any $x \in \R^2$, where the saturation level is symmetric and equal to $10$. We consider the measurement noise $w$ equal to $0.1\cos(10t)$ when $t \in [0,20]$, $0.01\cos(0.1t)$ when $t \in (20,40]$, $0.05\cos(100t)$ when $t \in (40,80]$ and  $0.1\cos(10t)$ when $t \in (80,100]$. 

\begin{figure*}
	\centering
	\includegraphics[trim= 1.1cm 9.7cm 1.1cm 10.0cm, clip, width=0.9\linewidth]{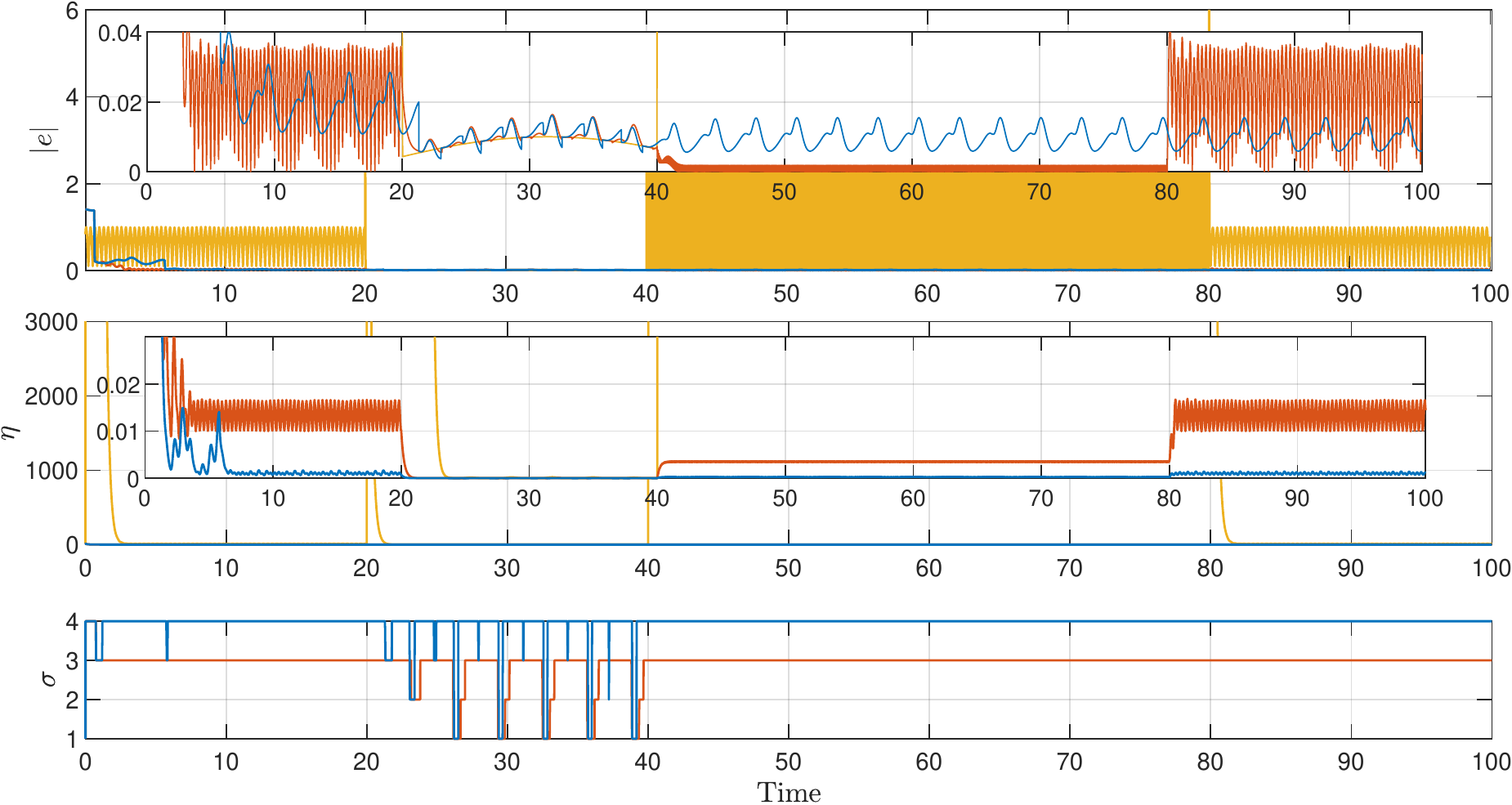}
	\caption{\normalfont{Norm of the estimation error $|e|$ (top figure), $\eta$ (middle figure) and $\sigma$ (bottom figure). Nominal (yellow), without resets (red), with resets (blue).
	}}
	\label{FIG:estimationErrorAndSigma_VanDerPol}
\end{figure*}

We design a nominal high-gain observer for system \eqref{eq:VanDerPolSystem} 
\begin{equation}
	\begin{aligned}
		\dot{\hat{x}}_{1} &= A\hat{x}_{1} + B\varphi(\hat{x}_{1}) + L_1 (y -\hat{y}_{1})\\
		\hat{y}_{1} &= C\hat{x}_{1}
	\end{aligned}
	\label{eq:VanDerPolObserver}
\end{equation}
where $\hat{x}_{1}$ is the state estimate, $\hat{y}_{1}$ is the estimated output and $L_1 \in \R^{2 \times 1}$ is the output injection gain, which is defined as $L_1 :=H_1D$, where $D \in \R^{2 \times 1}$, $H_{1} = \textnormal{diag}(h_{1}, h_{1}^2) \in \R^{2 \times 2}$, with $h_{1} \in \R_{> 0}$ the high-gain design parameter.
To satisfy Assumption~\ref{NominalAssumption}, 
$D \in \R^{2 \times 1}$ is selected such that the matrix $A-DC$ is Hurwitz and the parameter $h_{1}$ is taken sufficiently large, i.e., $h_{1} \geq h_1^\star$, where $h_1^\star$ is equal to $2\lambda_{\max}(P)K$, where $P \in \R^{2\times 2}$ is the solution of the Lyapunov equation $P(A-DC) + (A-DC)^\top P = -I_2$ and $K = 58.25$ is the Lipschitz constant of the function $\varphi$. 
We select $D$ such that the eigenvalues of $A-DC$ are equal to $-1$ and $-2$ and we obtain $D= [3, 2]$, while the parameter $h_{1}$ is selected equal to $200 > h_1^\star = 152.50$. 
With this choice of $h_{1}$, Assumption~\ref{NominalAssumption} is satisfied with a quadratic Lyapunov function and $\alpha = 53.28$. 

%
With the aim of improving the estimation performance, we consider $N = 4$ additional modes, with the same structure as the nominal one in \eqref{eq:VanDerPolObserver}. The only difference is the output injection gain $L_k \in \R^{2 \times 1}$, which is defined as $L_k := H_kD$, with $H_k  = \textnormal{diag}(h_{k}, h_{k}^2) \in \R^{2 \times 2}$, with $k \in \{2,\dots,5\}$. We select $h_2 = 20$, $h_3 = 1$, $h_4 =0$ and $h_5 =-1$. Note that $h_k \leq h_1^\star$, for all $k \in \{2,\dots,5\}$. Therefore,  we have no guarantees that these modes satisfy Assumption~\ref{NominalAssumption}, and consequently, that they converge. Simulations suggest that the modes with $L_2$ and $L_3$ converge, while 
the ones with $L_4$ and $L_5$ do not. 
Note that, for the reason given after Assumption~\ref{ASS:ass1}, the latter is satisfied.


We simulate the proposed estimation technique considering the initial conditions $x(0,0) = (1,1)$, $\hat{x}_k(0,0) = (0,0)$, $\eta_k(0,0) = 10$ for all $k \in \{1,\dots, 5\}$ and $\sigma(0,0) = 1$. 
 Both cases, without and with resets, are simulated 
with 
$\nu = 5$, $\lambda_1 =1$, $\lambda_2 = 1$ and $\varepsilon = 0.9$.  Note that the condition $\nu \in (0, \alpha]$ is satisfied. 

The norm of the nominal estimation error, namely $|e_1|$, as well as $|e_\sigma|$, obtained with or without resets, 
 are shown in Fig.~\ref{FIG:estimationErrorAndSigma_VanDerPol}, 
%
together with the nominal monitoring variable $\eta_1$ and 
the monitoring variables $\eta_{\sigma}$ obtained both in the case without and with resets. 
Fig.~\ref{FIG:estimationErrorAndSigma_VanDerPol} shows that both solutions (without resets and with resets) improve the estimation performance compared to the nominal one. 
Moreover, in this example, at the end of the simulation, it is clear that the resets reduce the norm of the estimation error and the corresponding monitoring variable, thereby further improving the estimation performance of the nominal observer. 
The last plot in Fig.~\ref{FIG:estimationErrorAndSigma_VanDerPol} represents $\sigma$ and indicates which mode is selected at every time instant both in the case without and with resets. Interestingly, when the resets are considered, the fourth mode (with $L_4 = 0$), that is not converging, appears to be selected.  

\section{Conclusion} \label{Conclusions}
We have presented a novel hybrid multi-observer that aims at improving the state estimation performance of a given nominal nonlinear observer. Each additional mode of the multi-observer differs from the nominal one only in its output injection gain, that can be freely selected as no convergence property is required for these modes. Inspired by supervisory control and observer approaches, we have designed a switching criterion, based on monitoring variables, that selects one mode at any time instant. We have proved an input-to-state stability property of the estimation error
and we have shown the efficiency of the proposed approach in improving the estimation performance in a numerical example. In future work, we will investigate analytical conditions to prove the strict estimation performance improvement. Moreover, it will also be interesting to exploit the proposed scheme for observer-based control.

\appendix
\noindent\textbf{Proof of Theorem~\ref{Prop:LyapunovSolutionProposition}.}
This proof relies on Proposition~\ref{THM:LyapunovTheorem}.
Consider the Lyapunov function $U$ in Proposition~\ref{THM:LyapunovTheorem}. From item (\textit{ii}) of Proposition~\ref{THM:LyapunovTheorem}, we have that for any $q \in \mathcal{C}$, $u \in \mathcal{U}$, $v \in \mathcal{V}$ and $w \in \mathcal{W}$ such that $F(q,u,v,w) \in T_{\mathcal{C}}(q)$,
\begin{equation}
	\begin{aligned}
		U^\circ(q,F(q,u,v,w)) \leq &-\alpha_0 U(q) + \phi_1(|v|)+\phi_2(|w|).
	\end{aligned}
	\label{eq:LyapunovFlowFromTh1}
\end{equation} 
We follow similar steps as in  \cite[proof of Theorem 3.18]{goebel2012hybrid}. 
Let $u \in \mathcal{U}$, $v \in \mathcal{V}$, $w \in \mathcal{W}$ and $q$ be a solution to system \eqref{eq:HybridSystemGeneral}-\eqref{eq:jumpSet}.
Pick any $(t,j) \in \dom q$ and let $0 = t_0 \leq t_1 \leq \dots \leq t_{j+1} = t$ satisfy $\dom  q \cap ([0,t]\times \{0,1,\dots,j\}) = \bigcup_{i=0}^j [t_i, t_{i+1}] \times \{i\}$. For each $i \in \{0,\dots,j\}$ and almost all $s \in [t_i, t_{i+1}]$, $q(s,i) \in \mathcal{C}$ and $\frac{d}{ds} q(s,i) \in F(q(s,i),u(s,i),v(s,i),w(s,i)) \cap T_\mathcal{C}(q(s,i))$, similarly to \cite[Lemma 5]{postoyan2014framework}. 
Hence, \eqref{eq:LyapunovFlowFromTh1} implies that, for all $i \in \{0, \dots j\}$ and almost all $s \in [t_i, t_{i+1}]$, 
\begin{equation}
	\begin{array}{l}
	U^\circ\left(q(s,i),\frac{d}{ds} q(s,i)\right) \\[0.5em]
\qquad	\leq -\alpha_0 U(q(s,i)) + \phi_1(\norm{v}_{[0,s]})+\phi_2(\norm{w}_{[0,s]}).
	\end{array}
	\label{eq:LyapunovFlowISSimplicitSolutions}
\end{equation}
In view of \cite{teel2000assigning}, we have that,  for almost all $s\in[t_i,t_{i+1}]$, 
\begin{equation}
	\frac{d}{ds}U(q(s,i)) \leq U^\circ\left(q(s,i),\frac{d}{ds} q(s,i)\right).
	\label{eq:LyapunovFlowClarkeUpperbound}
\end{equation}
From \eqref{eq:LyapunovFlowISSimplicitSolutions} and \eqref{eq:LyapunovFlowClarkeUpperbound}, for each $i \in \{0, \dots, j\}$ and for almost all $s \in [t_i, t_{i+1}]$, 
\begin{equation}
	\begin{array}{l}
	\displaystyle	\frac{d}{ds}U(q(s,i)) \\[0.5em] 
		\qquad  \leq -\alpha_0 U(q(s,i)) + \phi_1(\norm{v}_{[0,s]})+\phi_2(\norm{w}_{[0,s]}).
	\end{array}
	\label{eq:LyapunovFlowsSolutionNoClarke}
\end{equation}
Using \cite[Theorem III.16.2]{szarski1965differential}, from \eqref{eq:LyapunovFlowsSolutionNoClarke} we obtain that for almost all $s \in [t_i, t_{i+1}]$, for all $i \in \{0,\dots,j\}$, 
		$
		U(q(s,i))  \leq e^{ -\alpha_0 (s-t_i)}U(q(t_i,i))  +\int_{t_i}^{s}e^{-\alpha_0(s-\tau)}(\phi_1(\norm{v}_{[0,\tau]})+\phi_2(\norm{w}_{[0,\tau]}))d\tau  
		 \leq e^{ -\alpha_0 (s-t_i)}U(q(t_i,i))  
		 + \frac{1}{\alpha_0}(1- e^{-\alpha_0(s-t_i)}) (\phi_1(\norm{v}_{[0,s]})+\phi_2(\norm{w}_{[0,s]})).$
Thus,
\begin{equation}
	\begin{array}{l}
		U(q(t_{i+1},i)) \\[0.5em]
		\quad 
		\leq e^{ -\alpha_0 (t_{i+1}-t_i)}U(q(t_i,i)) 
		 \displaystyle + \frac{1}{\alpha_0}(1- e^{-\alpha_0(t_{i+1}-t_i)}) \\[0.5em] \qquad\Big(\phi_1(\norm{v}_{[0,t_{i+1}]})+\phi_2(\norm{w}_{[0,t_{i+1}]})\Big).
	\end{array}
	\label{eq:LyapunovFlowSolutuionKLbound2OPT2}
\end{equation}
On the other hand, from item \textit{(iii)} of Proposition~\ref{THM:LyapunovTheorem}, for each $i\in \{1,\dots,j\}$,
$	U(q(t_i,i)) - U(q(t_i,i-1)) \leq 0.$ 
From the last inequality and \eqref{eq:LyapunovFlowSolutuionKLbound2OPT2}, 
we deduce that for any $(t,j) \in \dom q$,
$	U(q(t,j)) \leq e^{- \alpha_0 t}U(q(0,0)) + \frac{1}{\alpha_0}(1- e^{-\alpha_0 t})  (\phi_1(\norm{v}_{[0,t]})+\phi_2(\norm{w}_{[0,t]})) \leq e^{- \alpha_0 t}U(q(0,0)) + \frac{1}{\alpha_0}(\phi_1(\norm{v}_{[0,t]})+\phi_2(\norm{w}_{[0,t]})) \leq e^{- \alpha_0 t}U(q(0,0)) + \frac{1}{\alpha_0}(\phi_1(\norm{v}_{[0,t]} + \norm{w}_{[0,t]})+\phi_2(\norm{v}_{[0,t]} + \norm{w}_{[0,t]})) =  e^{- \alpha_0 t}U(q(0,0)) + \tilde{\gamma}_U(\norm{v}_{[0,t]} + \norm{w}_{[0,t]})$, where $\tilde{\gamma}_U(s) = \frac{1}{\alpha_0} ( \phi_1(s) + \phi_2(s)) \in \Kinf$. 
%
Using item (\textit{i}) of Proposition~\ref{THM:LyapunovTheorem}, we obtain, for any $(t,j) \in \dom q$, 
$|(e_1(t,j), \eta_1(t,j), e_{\sigma} (t,j), \eta_{\sigma}(t,j))| ) \leq \underline{\alpha}_U^{-1}(e^{- \alpha_0 t}\overline{\alpha}_U(|(e(0,0), \eta(0,0))|)) + \tilde{\gamma}_U(\norm{v}_{[0,t]} + \norm{w}_{[0,t]}))$. Since for any $\alpha \in \Kinf$ we have $\alpha(s_1 + s_2) \leq \alpha(2s_1) + \alpha(2 s_2)$ for all $s_1 \geq 0, s_2 \geq 0$, we obtain 
 \eqref{eq:LyapunovSolutionTheoremEquation} where $\beta_U(r,s):= \underline{\alpha}_U^{-1}(2e^{- \alpha_0 s}\overline{\alpha}_U(r)) \in \KL$ and $\gamma_U(r) := \underline{\alpha}_U^{-1}(2\tilde{\gamma}_U(r)) \in \Kinf$ for all $r,s \geq 0$. 
\hfill $\blacksquare$

\bibliography{bibliography}

\begin{thebibliography}{10}

\bibitem{bernard2022observer}
P.~Bernard, V.~Andrieu, and D.~Astolfi, ``Observer design for continuous-time
  dynamical systems,'' {\em Annual Reviews in Control}, 2022.

\bibitem{kalman1961new}
R.~E. Kalman and R.~S. Bucy, ``New results in linear filtering and prediction
  theory,'' {\em Journal of Basic Engineering}, vol.~83, no.~1, pp.~95--108,
  1961.

\bibitem{helton1999extending}
J.~W. Helton and M.~R. James, {\em Extending $H_\infty$ control to nonlinear
  systems: control of nonlinear systems to achieve performance objectives}.
\newblock SIAM, 1999.

\bibitem{li2015finite}
Y.~Li and R.~G. Sanfelice, ``A finite-time convergent observer with robustness
  to piecewise-constant measurement noise,'' {\em Automatica}, vol.~57,
  pp.~222--230, 2015.

\bibitem{rios2018hybrid}
H.~R{\'\i}os and A.~R. Teel, ``A hybrid fixed-time observer for state
  estimation of linear systems,'' {\em Automatica}, vol.~87, pp.~103--112,
  2018.

\bibitem{alessandri2022hysteresis}
A.~Alessandri and R.~G. Sanfelice, ``Hysteresis-based switching observers for
  linear systems using quadratic boundedness,'' {\em Automatica}, vol.~136,
  p.~109982, 2022.

\bibitem{astolfi2015high}
D.~Astolfi and L.~Marconi, ``A high-gain nonlinear observer with limited gain
  power,'' {\em IEEE Transactions on Automatic Control}, vol.~60, no.~11,
  pp.~3059--3064, 2015.

\bibitem{astolfi2018low}
D.~Astolfi, L.~Marconi, L.~Praly, and A.~R. Teel, ``Low-power peaking-free
  high-gain observers,'' {\em Automatica}, vol.~98, pp.~169--179, 2018.

\bibitem{esfandiari2019bank}
K.~Esfandiari and M.~Shakarami, ``Bank of high-gain observers in output
  feedback control: Robustness analysis against measurement noise,'' {\em IEEE
  Transactions on Systems, Man, and Cybernetics: Systems}, vol.~51, no.~4,
  pp.~2476--2487, 2019.

\bibitem{bernat2015multi}
J.~Bernat and S.~Stepien, ``Multi-modelling as new estimation schema for
  high-gain observers,'' {\em International Journal of Control}, vol.~88,
  no.~6, pp.~1209--1222, 2015.

\bibitem{mousavi2021low}
S.~Mousavi and M.~Guay, ``A low-power multi high-gain observer design for state
  estimation in nonlinear systems,'' {\em IEEE Conference on Decision and
  Control, \textnormal{Austin, USA}}, pp.~5435--5440, 2021.

\bibitem{ahrens2009high}
J.~H. Ahrens and H.~K. Khalil, ``High-gain observers in the presence of
  measurement noise: A switched-gain approach,'' {\em Automatica}, vol.~45,
  no.~4, pp.~936--943, 2009.

\bibitem{astolfi2021stubborn}
D.~Astolfi, A.~Alessandri, and L.~Zaccarian, ``Stubborn and dead-zone redesign
  for nonlinear observers and filters,'' {\em IEEE Transactions on Automatic
  Control}, vol.~66, no.~2, pp.~667--682, 2021.

\bibitem{battilotti2021performance}
S.~Battilotti, ``Performance optimization via sequential processing for
  nonlinear state estimation of noisy systems,'' {\em IEEE Transactions on
  Automatic Control}, 2021.

\bibitem{astolfi2019uniting}
D.~Astolfi, R.~Postoyan, and D.~Ne{\v{s}}i{\'c}, ``Uniting observers,'' {\em
  IEEE Transactions on Automatic Control}, vol.~65, no.~7, pp.~2867--2882,
  2019.

\bibitem{chong2015parameter}
M.~S. Chong, D.~Ne{\v{s}}i{\'c}, R.~Postoyan, and L.~Kuhlmann, ``Parameter and
  state estimation of nonlinear systems using a multi-observer under the
  supervisory framework,'' {\em IEEE Transactions on Automatic Control},
  vol.~60, no.~9, pp.~2336--2349, 2015.

\bibitem{aguiar2008identification}
A.~P. Aguiar, V.~Hassani, A.~M. Pascoal, and M.~Athans, ``Identification and
  convergence analysis of a class of continuous-time multiple-model adaptive
  estimators,'' {\em IFAC Proceedings Volumes}, vol.~41, no.~2, pp.~8605--8610,
  2008.

\bibitem{aguiar2007convergence}
A.~P. Aguiar, M.~Athans, and A.~M. Pascoal, ``Convergence properties of a
  continuous-time multiple-model adaptive estimator,'' {\em European Control
  Conference, \textnormal{Kos, Greece}}, pp.~1530--1536, 2007.

\bibitem{shim2015nonlinear}
H.~Shim and D.~Liberzon, ``Nonlinear observers robust to measurement
  disturbances in an {ISS} sense,'' {\em IEEE Transactions on Automatic
  Control}, vol.~61, no.~1, pp.~48--61, 2015.

\bibitem{hespanha2003hysteresis}
J.~P. Hespanha, D.~Liberzon, and A.~S. Morse, ``Hysteresis-based switching
  algorithms for supervisory control of uncertain systems,'' {\em Automatica},
  vol.~39, no.~2, pp.~263--272, 2003.

\bibitem{morse1996supervisory}
A.~S. Morse, ``Supervisory control of families of linear set-point
  controllers-part i. exact matching,'' {\em IEEE Transactions on Automatic
  Control}, vol.~41, no.~10, pp.~1413--1431, 1996.

\bibitem{hespanha2001multiple}
J.~Hespanha, D.~Liberzon, A.~Stephen~Morse, B.~D. Anderson, T.~S. Brinsmead,
  and F.~De~Bruyne, ``Multiple model adaptive control. part 2: switching,''
  {\em International Journal of Robust and Nonlinear Control}, vol.~11, no.~5,
  pp.~479--496, 2001.

\bibitem{goebel2012hybrid}
R.~Goebel, R.~G. Sanfelice, and A.~R. Teel, {\em Hybrid Dynamical Systems:
  Modeling, Stability, and Robustness}.
\newblock New Jersey, USA: Princeton University Press, 2012.

\bibitem{heemels2021hybrid}
W.~P. M.~H. Heemels, P.~Bernard, K.~J.~A. Scheres, R.~Postoyan, and R.~G.
  Sanfelice, ``Hybrid systems with continuous-time inputs: Subtleties in
  solution concepts and existence properties,'' {\em IEEE Conference on
  Decision and Control, \textnormal{Austin, USA}}, pp.~5361--5366, 2021.

\bibitem{clarke1990optimization}
F.~H. Clarke, {\em Optimization and Nonsmooth Analysis}.
\newblock Philadelphia, USA: Classic in Applied Mathematics vol. 5, SIAM, 1990.

\bibitem{astolfi2021constrained}
D.~Astolfi, P.~Bernard, R.~Postoyan, and L.~Marconi, ``Constrained state
  estimation for nonlinear systems: a redesign approach based on convexity,''
  {\em IEEE Transactions on Automatic Control}, vol.~67, no.~2, pp.~824--839,
  2021.

\bibitem{willems2004deterministic}
J.~C. Willems, ``Deterministic least squares filtering,'' {\em Journal of
  Econometrics}, vol.~118, no.~1-2, pp.~341--373, 2004.

\bibitem{na2017adaptive}
J.~Na, G.~Herrmann, and K.~G. Vamvoudakis, ``Adaptive optimal observer design
  via approximate dynamic programming,'' {\em American Control Conference,
  \textnormal{Seattle, USA}}, pp.~3288--3293, 2017.

\bibitem{khalil2014high}
H.~K. Khalil and L.~Praly, ``High-gain observers in nonlinear feedback
  control,'' {\em International Journal of Robust and Nonlinear Control},
  vol.~24, no.~6, pp.~993--1015, 2014.

\bibitem{cai2007smooth}
C.~Cai, A.~R. Teel, and R.~Goebel, ``Smooth lyapunov functions for hybrid
  systems-part i: Existence is equivalent to robustness,'' {\em IEEE
  Transactions on Automatic Control}, vol.~52, no.~7, pp.~1264--1277, 2007.

\bibitem{sontag2008input}
E.~D. Sontag, ``Input to state stability: Basic concepts and results,'' in {\em
  Nonlinear and optimal control theory}, pp.~163--220, Springer, 2008.

\bibitem{postoyan2014framework}
R.~Postoyan, P.~Tabuada, D.~Ne{\v{s}}i{\'c}, and A.~Anta, ``A framework for the
  event-triggered stabilization of nonlinear systems,'' {\em IEEE Transactions
  on Automatic Control}, vol.~60, no.~4, pp.~982--996, 2014.

\bibitem{teel2000assigning}
A.~R. Teel and L.~Praly, ``On assigning the derivative of a disturbance
  attenuation control lyapunov function,'' {\em Mathematics of Control, Signals
  and Systems}, vol.~13, no.~2, pp.~95--124, 2000.

\bibitem{szarski1965differential}
J.~Szarski, {\em Differential inequalities}.
\newblock Polish Scientific Publisher, Warsaw, 1965.

\end{thebibliography}

\end{document}